\documentclass[aps,twocolumn]{revtex4}
\begin{document}
\title{Physical origin of the power-law tailed statistical distributions}
\author{\footnotesize{G. Kaniadakis}}
\address{Department of Applied Science and Technology, Politecnico di Torino, \\
Corso Duca degli Abruzzi 24, 10129 Torino, Italy\\
giorgio.kaniadakis@polito.it}

\begin {abstract}
Starting from the BBGKY hierarchy, describing the kinetics of
nonlinear particle system, we obtain the relevant entropy and
stationary distribution function. Subsequently, by employing the
Lorentz transformations we propose the relativistic generalization
of the exponential and logarithmic functions. The related particle
distribution and entropy represents the relativistic extension of
the classical Maxwell-Boltzmann distribution and of the Boltzmann
entropy respectively and define the statistical mechanics presented
in [Phys. Rev. E {\bf 66}, 056125 (2002)] and [Phys. Rev. E {\bf
72}, 036108 (2005). The achievements of the present effort, support
the idea that the experimentally observed power law tailed
statistical distributions in plasma physics, are enforced by the
relativistic microscopic particle dynamics.

\end {abstract}

\maketitle


\section{Introduction}

In plasma physics, the power-law tails in the particle population,
has been systematically observed in the last fifty years. For
instance the cosmic ray spectrum
\begin{equation}
f_i\propto \chi(\beta E_i-\beta \mu) \ \ , \label{A1}
\end{equation}
which extends over 13 decades in energy, from $10^8$ eV to $10^{20}$
eV, and spans 33 decades in particle flux, from $10^{-29}$ to $10^4$
units, obeys the Boltzmann law of classical statistical mechanics
i.e.
\begin{equation}
\chi(x) {\atop\stackrel{\textstyle\sim}{\scriptstyle x\rightarrow
\,0}} \exp(-x) \ \ , \label{A2}
\end{equation}
for low energies, while for high energies this spectrum presents
power law fat tails i.e
\begin{equation}
\chi(x) {\atop\stackrel{\textstyle\sim}{\scriptstyle x\rightarrow
\,+\infty}} x^{-1/\kappa} \ \ , \label{A3}
\end{equation}
the spectral index $\kappa$ being close to 0.32-0.37.

The above spectrum was approached for the first time in 1968, by
using a different distribution from the Boltzmann one. In his
proposal Vasyliunas heuristically identified the function $\chi(x)$
with the Student distribution function which presents power-law
tails \cite{Vasyliunas}. In the last 40 years  a vast amount of
literature has been produced, regarding the so called kappa-plasmas,
based on the Vasyliunas distribution. Up to now, several physical
mechanisms have been explored in order to furnish theoretical
support to the experimentally observed power-law-tailed distribution
functions. However there is currently an intense debate regarding
the theoretical foundations of the non-Boltzmannian distributions.

In the last years, after noting that the power-law tails are placed
in the high energy region, and then regards relativistic particles,
the question has been posed whether the solution of the problem,
i.e. the theoretic determination of the function $\chi(x)$ and
consequently of the related distribution and entropy, can be
explained by invoking the basic principles of special relativity.

The present paper, deals with the relativistic
statistical theory \cite{PhA2001,PLA2001,PRE2002,PRE2005,PhA2006,EPJB2009,EPJA2009,EPL2010,PLA2011},
predicting for the function $\chi(x)$, the very simple form i.e.
\begin{equation}
\chi(x)=\exp_{\kappa}(-x) \ , \label{A4}
\end{equation}
with
\begin{equation}
\exp_{\kappa}(x)=(\sqrt{1+ \kappa^2 x^2}+\kappa x)^{1/\kappa}.
\label{A5}
\end{equation}
The parameter $\kappa<1$ is the reciprocal of light speed in a
dimensionless form while the function $\exp_{\kappa}(x)$ represents
the relativistic generalization of the ordinary exponential which
recovers in the classical limit $\kappa \rightarrow 0$.

In the last few years various authors have considered the
foundations of the statistical theory based on the distribution
function involving the generalized exponential $\exp_{\kappa}(x)$,
e.g. the H-theorem and the molecular chaos hypothesis
\cite{Silva06A,Silva06B}, the thermodynamic stability
\cite{Wada1,Wada2}, the Lesche stability
\cite{KSPA04,AKSJPA04,Naudts1,Naudts2}, the Legendre structure of
the ensued thermodynamics \cite{ScarfoneWada,Yamano},
the thermodynamics of non-equilibrium systems \cite{Lucia2010},
quantum versions of the theory \cite{AlianoKM2003,Santos2011a,Santos2011b},
the geometrical structure of the theory \cite{Pistone},
various mathematical aspects of the theory
\cite{KLS2004,KLS2005,KSphysA2003,Oikonomou2010,Stankovic2011,Tempesta2011}, etc.
On the other hand specific applications to physical systems have been
considered, e.g. the cosmic rays \cite{PRE2002}, relativistic
\cite{GuoRelativistic} and classical \cite{GuoClassic} plasmas  in
presence of external electromagnetic fields, the relaxation in
relativistic plasmas under wave-particle interactions
\cite{Lapenta,Lapenta2009},  anomalous diffusion
\cite{WadaScarfone2009,Wada2010}, non-linear kinetics \cite{KQSphysA2003,BiroK2006},
kinetics of interacting atoms and
photons \cite{Rossani}, particle kinetics in the presence of
temperature gradients \cite{GuoDuoTgradient},  particle systems in
external conservative force fields \cite{Silva2008}, stellar
distributions in astrophysics \cite{Carvalho,Carvalho2,Carvalho2010},
quark-gluon plasma formation \cite{Tewel}, quantum hadrodynamics models
\cite{Pereira}, the fracture propagation \cite{Fracture}, etc. Other
applications regard dynamical systems at the edge of chaos
\cite{Corradu,Tonelli,Celikoglu}, fractal systems \cite{Olemskoi},
field theories \cite{Olemskoi2010},
the random matrix theory \cite{AbulMagd,AbulMagd2009}, the error
theory \cite{WadaSuyari06}, the game theory \cite{Topsoe}, the
information theory \cite{WadaSuyari07}, etc. Also applications to
economic systems have been considered e.g. to study the personal
income distribution \cite{Clementi,Clementi2008,Clementi2009,Clementi2011}, to
model deterministic heterogeneity in tastes and product
differentiation \cite{Rajaon,Rajaon2008} etc.
Finally in \cite{KL2004}, some historical remarks, on the theories
dealing with power-law tailed distribution functions, are reported.

In the present contribution we reconsider critically the foundations
of the  statistical theory the generalized exponential (\ref{A5}).
Our main goal is to show that  (i) the function $\exp_{\kappa}(x)$,
can be obtained within the one-particle relativistic dynamics, in a
very simple and transparent way, starting directly from the Lorentz
transformations (ii) the standard principles of ordinary
relativistic kinetics, conduct unambiguously to the relativistic
generalization of the classical Boltzmann entropy and
Maxwell-Boltzmann distribution, without invoking any extra principle
or assumption.

\section{Kinetic Equation}

Let us consider the most general relativistic equation imposing the
particle conservation during the evolution of a many body system.
That equation is the first equation of the BBGKY hierarchy i.e.
\begin{equation}
p^{\,\nu}\partial_{\nu}f-m F^{\nu}\frac{\partial f}{\partial
p^{\,\nu}}= \!\int \!
\frac{d^3p'}{{p'}^{0}}\frac{d^3p_1}{p_1^{\,0}}\frac{d^3p'_1}{{p'}_{\!\!1}^{0}}
\,\,G \,\left[f'\!\otimes \!f'_1\!-\! f\!\otimes \! f_1\right],
\label{B1}
\end{equation}
and describes, through the one-particle correlation function or
distribution function $f=f(x,p)$, a relativistic particle system in
presence of an external force field. The streaming term as well as
the Lorentz invariant integrations in the collision integral, have
the standard forms of the relativistic kinetic theory
\cite{Degroot,Cercignani}.

The two particle correlation function \cite{KersonHuang}, here
denoted by $f\!\otimes \! f_1$, appearing in Eq. (\ref{B1}), at the
moment remains an unknown, two variable function. We recall that in
classical kinetics, the two-particle correlation function, according
to the molecular chaos hypothesis, is the ordinary product of $f$
and $f_1$ i.e. $f\!\otimes \! f _1= f\,f_1$. Therefore the
composition law $f\otimes f_1$ can be viewed as a relativistic
generalized product of $f$ and $f_1$, isomorphic to the ordinary
product.

Following standard lines of kinetic theory, we note that in
stationary conditions, the collision integral in Eq. (\ref{B1})
vanishes and then it follows that $f\!\otimes \! f_1= f'\!\otimes \!
f'_1 $. More in general it holds
\begin{equation}
L(f\!\otimes \! f_1)= L(f'\!\otimes \! f'_1) \ , \ \ \label{}
\end{equation}
$L(x)$ being an arbitrary function, and this relationship expresses
a conservation law for the particle system. On the other hand a
conservation law has form
\begin{equation}
\Lambda(f)+ \Lambda(f_{1}) =\Lambda(f^\prime) +\Lambda(f^\prime_{1})
\ , \ \ \label{B3}
\end{equation}
$\Lambda (f)$ being the collision invariant of the system. Therefore
we can pose
\begin{equation}
L(f\!\otimes\!f_1)= \Lambda(f)+\Lambda(f_1) \ . \ \ \label{B3a}
\end{equation}

From the definition of the correlation function and taking into
account of the indistinguishability of the particles, it follows that
$f\!\otimes \!1=1\otimes \!f=f$ and this implies that $\Lambda(1)=0$
and $L(f)=\Lambda(f)$. After taking into account the later
relationship, Eq. (\ref{B3a}) becomes
\begin{eqnarray}
\Lambda(f\!\otimes\!f_1)=  \Lambda(f)+\Lambda(f_1) \ \ . \label{B4a}
\end{eqnarray}
Consequently the collision invariant $\Lambda(f)$ permits us to
determine univocally the correlation function $f\!\otimes\!f_1$ as
follows
\begin{eqnarray}
f\!\otimes\!f_1= \Lambda^{-1} \Big ( \Lambda(f)+\Lambda(f_1)\Big ) \
\ . \label{B4}
\end{eqnarray}

In relativistic kinetics, the collision invariant $\Lambda(f)$,
unless an additive constant, is proportional to the microscopic
relativistic invariant $I$. Then we can pose
\begin{equation}
\Lambda(f)=- \beta I +\beta \mu \ , \label{B6}
\end{equation}
$\beta $ and $\beta \mu$ being two arbitrary constants. In presence
of an external electromagnetic field $A^{\nu}$, the more general
microscopic relativistic invariant $I$, has a form proportional to
\begin{equation}
I=\left(p^{\nu}+q A^{\nu}\!/c \right)\,U_{\nu}-mc^2 \ , \label{B7}
\end{equation}
$U_{\nu}$ being the hydrodynamic four-vector velocity with
$U^{\nu}U_{\nu}=c^2$ \cite{Degroot}. Finally, after inversion of Eq.
(\ref{B6}),  the stationary distribution is obtained as follows
\begin{equation}
f=\Lambda^{-1}\big(-\beta\,I+\beta \,\mu \,\big) \ . \label{B8}
\end{equation}

It is remarkable to note that Eq. (\ref{B6}) follows from the
variational equation
\begin{equation}
\frac{\delta}{\delta f}\,\,\Bigg[ \, S \,-\, \beta \!\!\int \!\!d^3p
\,\,I\,f\, \,+\, \beta \mu \!\!\int\!\! d^3p \,\, f \,\Bigg]=0 \ ,
\ \ \ \label{B9}
\end{equation}
where the functional $S$, unless an arbitrary additive constant, is
given by
\begin{equation}
S= - \int \!\!d^3p \,\left [\,\int \Lambda (f) \, df \,\right ]  \ ,
\ \ \ \ \ \ \ \ \ \ \label{B10}
\end{equation}
$\int \Lambda (f)\,\, df$ being the indefinite integral of $\Lambda
(f)$. The variational equation (\ref{B9}) represents  the maximum
entropy principle. The constants $\beta$ and $\beta \mu$ are the
Lagrange multipliers while the functional $S$, defined though Eq.
(\ref{B10}), is the system entropy.

We stress that the function $\Lambda (f)$ defines univocally both
the stationary distribution (\ref{B8}) and the entropy (\ref{B10})
of the system as well as the two-particle correlation function
(\ref{B4}). In classical statistical mechanics, it is well known
that $\Lambda(f)=\ln(f)$ so that the two particle correlation
function becomes $f\otimes f_1=f\,f_1$, while (\ref{B8}) and
(\ref{B10}) reduces to the exponential distribution and Boltzmann
entropy respectively.

In the next section, in order to develop a relativistic statistical
mechanics, we will determine the function $\Lambda(f)$ within the
special relativity, starting from the Lorentz transformations.

\section{Lorentz Transformations}

In the present section we will show that the function
$\Lambda^{-1}(f)$ emerges within the special relativity as the
relativistic generalization of the ordinary exponential of classical
physics.

Let us consider in the one-dimension frame $\cal S$ two identical
particles $A$ and $B$, of rest mass $m$. We suppose that the two
particles move with velocity $v_{\scriptscriptstyle A}$ and
$v_{\scriptscriptstyle B}$ respectively. The momenta of the two
particles are indicated with $p_{\scriptscriptstyle
A}=p\,(v{\scriptscriptstyle A})$ and $p_{\scriptscriptstyle
B}=p\,(v_{\scriptscriptstyle B})$, while their energies are
indicated with ${E}_{\scriptscriptstyle
A}={E}\,(v_{\scriptscriptstyle A})$ and ${E}_{\scriptscriptstyle
B}={E}\,(v_{\scriptscriptstyle B})$ respectively.

In classical physics, in the rest frame ${\cal S}'$ of particle $B$,
the momentum of the particle $B$ is $p'_{\scriptscriptstyle B}=0$
while the momentum of the particle $A$ is given by the Galilei
transformation formula
\begin{eqnarray}
p'_{\scriptscriptstyle A}= p_{\scriptscriptstyle
A}-p_{\scriptscriptstyle B} \ . \label{DA1}
\end{eqnarray}
After introducing in place of $p$ the dimensionless momentum
$q=p/p_*$, we note that the exponential function $\exp(q)$, permits
us to write the Galilei additivity law (\ref{DA1}), in the following
factorized form
\begin{eqnarray}
\exp(q'_{\scriptscriptstyle A})= \exp(q_{\scriptscriptstyle
A})\exp(-q_{\scriptscriptstyle B}) \ . \label{DA2}
\end{eqnarray}

The Galilei relativity principle, imposes the equivalence of all the
inertial frames. According to this principle, the inverse Galilei
transformation must have the same structure of the direct
transformation (\ref{DA2}) except for the substitutions
$q'_{\scriptscriptstyle A} \leftrightarrow q_{\scriptscriptstyle A}$
and $q_{\scriptscriptstyle B} \rightarrow - q_{\scriptscriptstyle
B}$. This requirement is satisfied thanks to the following property
of the exponential function
\begin{eqnarray}
\exp(x)\exp(-x)= 1 \ \ . \label{DA3}
\end{eqnarray}

We consider now the two particles in the rest frame ${\cal S}'$ of
particle $B$, within the special relativity. The velocity, momentum
and energy of the particle $B$ are $v'_{\scriptscriptstyle B}=0$,
$p'_{\scriptscriptstyle B}=0$ and $E'_{\scriptscriptstyle B}=mc^2$
respectively. In ${\cal S}'$ the velocity of particle $A$ is given
by the formula
\begin{eqnarray}
v'_{\scriptscriptstyle A}=\frac{v_{\scriptscriptstyle
A}-v_{\scriptscriptstyle B}}{1-v_{\scriptscriptstyle A}
v_{\scriptscriptstyle B}/c^2}\ \ . \label{DA4}
\end{eqnarray}
defining the relativistic velocity composition law, which follows
directly from the kinematic Lorentz transformations. In the same
frame ${\cal S}'$ the momentum and energy of particle $A$ are given
by the dynamic Lorentz transformations
\begin{eqnarray}
&&p'_{\scriptscriptstyle A}=\gamma( v_{\scriptscriptstyle B})
p_{\scriptscriptstyle A}-c^{-2}v_{\scriptscriptstyle B}\gamma(
v_{\scriptscriptstyle B})E_{\scriptscriptstyle A} \ ,
\label{D1} \\
&&E'_{\scriptscriptstyle A}=\gamma( v_{\scriptscriptstyle B})
E_{\scriptscriptstyle A}-v_{\scriptscriptstyle B}\gamma(
v_{\scriptscriptstyle B})p_{\scriptscriptstyle A} \ , \label{D2}
\end{eqnarray}
$\gamma(v_{\scriptscriptstyle B})=(1-v_{\scriptscriptstyle
B}^2/c^2)^{-1/2}$ being the Lorentz factor. After taking into
account the expression of the momentum $p_{\scriptscriptstyle
B}=mv_{\scriptscriptstyle B}\gamma(v_{\scriptscriptstyle B})$ and of
the energy $E_{\scriptscriptstyle
B}=mc^2\gamma(v_{\scriptscriptstyle B})$ of the particle $B$ the
latter transformations become
\begin{eqnarray}
&&p'_{\scriptscriptstyle A}= p_{\scriptscriptstyle
A}E_{\scriptscriptstyle B}/mc^2-E_{\scriptscriptstyle
A}p_{\scriptscriptstyle B}/mc^2 \ ,
\label{DA5} \\
&&E'_{\scriptscriptstyle A}= E_{\scriptscriptstyle
A}E_{\scriptscriptstyle B}/mc^2-p_{\scriptscriptstyle
A}p_{\scriptscriptstyle B}/m \ , \label{DA6}
\end{eqnarray}

Let us introduce in place of the dimensional variables $(v,p,E)$ the
dimensionless variables $(u,q,{\cal E})$ through
\begin{eqnarray}
\frac{v}{u}=\frac{p}{mq}=\sqrt{\frac{E}{m{\cal E}}}=|\kappa| c
=v_*<c \ . \label{DA7}
\end{eqnarray}

From its definition, it follows that $\kappa$ can be viewed
as the Einstein $\beta$ factor related to the velocity $v_*$.  The
condition $v_*<c$, implies that $-1<\kappa <+1$.

The dynamic Lorentz transformations for the dimensionless momentum
and energy variables become
\begin{eqnarray}
&&q'_{\scriptscriptstyle A}=\kappa^2 q_{\scriptscriptstyle A} {\cal
E}_{\scriptscriptstyle B} - \kappa^2 q_{\scriptscriptstyle B} {\cal
E}_{\scriptscriptstyle A} \ ,
\label{D3} \\
&&{\cal E}'_{\scriptscriptstyle A}=\kappa^2 {\cal
E}_{\scriptscriptstyle A} {\cal E}_{\scriptscriptstyle B} -
q_{\scriptscriptstyle A} q_{\scriptscriptstyle B} \ , \label{D4}
\end{eqnarray}
while the classical limit $c\rightarrow \infty$ is replaced by the
limit $\kappa\rightarrow 0$.

For a particle at rest it results $E(0)=m\,c^2$ and then ${\cal E}(0)=1/\kappa^2$.
Then $1/\kappa^2$ represents the dimensionless rest energy of the
particle. Alternatively $1/\kappa$ can be viewed as the refractive
index of a medium in which the light speed is $v_*$.

From the Lorentz invariance it follows
easily the energy-momentum dispersion relation
\begin{eqnarray}
\kappa^4{\cal E}^2-\kappa^2q^2=1 \ . \label{DA8}
\end{eqnarray}
After expressing in the right hand side of Eq. (\ref{D3}), the energy in
terms of the momentum ${\cal E}=\sqrt{1+\kappa^2q^2}/\kappa^2 $,  we
obtain the momentum relativistic additivity law as follows
\begin{eqnarray}
q'_{\scriptscriptstyle A}=q_{\scriptscriptstyle A}
\sqrt{1+\kappa^2q_{\scriptscriptstyle B}^2} - q_{\scriptscriptstyle
B} \sqrt{1+\kappa^2q_{\scriptscriptstyle A}^2} \ , \label{DA9}
\end{eqnarray}
which in the classical limit $\kappa \rightarrow 0$ reproduces the
ordinary additivity law (\ref{DA1}), of classical physics.

The Galilei relativity principle, holding both in classical physics
and in special relativity, imposes the equivalence of all the
inertial frames. According to this principle, the inverse
transformation of (\ref{DA9}) must have the same structure of the
direct transformations (\ref{DA9}) except for the substitutions
$q'_{\scriptscriptstyle A} \leftrightarrow q_{\scriptscriptstyle A}$
and $q_{\scriptscriptstyle B} \rightarrow - q_{\scriptscriptstyle
B}$ i.e.

\begin{eqnarray}
q_{\scriptscriptstyle A}=q'_{\scriptscriptstyle A}
\sqrt{1+\kappa^2{q}_{\scriptscriptstyle B}^2} +
q_{\scriptscriptstyle B} \sqrt{1+\kappa^2 {q_{\scriptscriptstyle
A}'}^2}  \ , \label{DA10}
\end{eqnarray}
It is easy to verify that (\ref{DA10}) follows directly from
(\ref{DA9}) and viceversa.

The Lorentz transformation for the relativistic momenta (\ref{DA9}),
representing the momenta additivity law in special relativity, has
the important feature that the contributions of the two particles,
appearing in the right hand side of the equation, are
not factorized. Spontaneously the question emerges at this point,
whether new variables exist, able to factorize the contribution of
the two particles in the right hand side of the relativistic additivity
law (\ref{DA9}).

It is easy to verify that the new variable is given by function
\begin{eqnarray}
\exp_{\kappa}\!\left( q\right) = \left( \sqrt{1+\kappa^2 q^2}
+\kappa q \right )^{1/\kappa} \ , \label{DA11}
\end{eqnarray}
so that the Lorentz transformation (\ref{DA9}) assumes the following
factorized form
\begin{eqnarray}
\exp_{\kappa}(q'_{\scriptscriptstyle A})=\exp_{\kappa}( -
q_{\scriptscriptstyle B}) \, \exp_{\kappa}( q_{\scriptscriptstyle
A}) \ \ . \label{DA12}
\end{eqnarray}
On the other hand Galilei relativity principle imposes, for the inverse
Lorentz transformation, the following factorized form
\begin{eqnarray}
\exp_{\kappa}(q_{\scriptscriptstyle A})=\exp_{\kappa}(
q_{\scriptscriptstyle B}) \exp_{\kappa}(q'_{\scriptscriptstyle A}) \
\ . \label{DA13}
\end{eqnarray}
By comparison of the above direct and inverse Lorentz
transformations, it obtains the property
\begin{eqnarray}
\exp_{\kappa}(q) \exp_{\kappa}(- q)=1 \ \ , \label{DA14}
\end{eqnarray}
which can be verified easily, by direct inspection of the definition
(\ref{DA11}).

It is remarkable that the function $\exp_{\kappa}(q)$ emerges in
one-particle special relativity as the variable, able to factorize
the momentum Lorentz transformation, and represents the relativistic
generalization of the ordinary exponential which factorize the
momentum Galilei transformation in classical physics. Clearly in the
classical limit $\kappa \rightarrow 0$, $\exp_{\kappa}(q)$ reduces
to $\exp(q)$.

The inverse function of $\exp_{\kappa}(q)$ indicated by
$\ln_{\kappa}(q)$ and defined through
$\ln_{\kappa}(\exp_{\kappa}(q))=\exp_{\kappa}(\ln_{\kappa}(q))=1$,
represents the relativistic generalization of the ordinary
logarithm, which recovers in the classical limit $\kappa \rightarrow
0$, and is given by
\begin{eqnarray}
\ln_{\kappa}\!\left( q\right) =
\frac{q^{\kappa}-q^{-\kappa}}{2\kappa} \ . \label{DA15}
\end{eqnarray}
The property (\ref{DA14}) of $\exp_{\kappa}(q)$, enforced by the
Galilei relativity principle, transforms into the following property
of $\ln_{\kappa}(x)$
\begin{eqnarray}
\ln_{\kappa}(1/x)= - \ln_{\kappa}(x) \ \ , \label{DA16}
\end{eqnarray}
holding also for the ordinary logarithm of classical physics.

\section{Relativistic Statistical Mechanics}

In the previous section it has been shown that the functions
$\exp_{\kappa}(q)$ and $\ln_{\kappa}(q)$ emerge as the relativistic
generalizations of the ordinary exponential and logarithm functions
of classical physics. Therefore in the following we pose
\begin{eqnarray}
&&\Lambda (x)= \ln_{\kappa} (x) \ \ , \label{CA1} \\
&&\Lambda^{-1} (x)= \exp_{\kappa} (x) \ \ . \label{CA2}
\end{eqnarray}
After taking into account of the property
\begin{eqnarray}
\frac{d}{dx}\,\,x\,\ln_{\kappa}(x)=\frac{1}{\gamma}
\,\ln_{\kappa}(\epsilon x) \ \ , \label{CA3}
\end{eqnarray}
with
\begin{eqnarray}
&&\gamma=\frac{1}{\sqrt{1-\kappa^2}} \ ,  \label{gamma} \\
&&\epsilon=\exp_{\kappa}(\gamma)  , \  \label{EPS}
\end{eqnarray}
the entropy (\ref{B10}) simplifies as
\begin{equation}
S= - \,\gamma\! \int \!d^3p \,\, f \,\ln_{\kappa} (f/\epsilon) \ .
\label{CA6}
\end{equation}

It is worth stressing that the latter relationship defines the
relativistic entropy, as proportional to the mean value of
$-\ln_{\kappa} (f/\epsilon)$, like in the case of classical
statistical mechanics where the Boltzmann entropy, $S= - \int d^3p
\,\, f \,\ln\,(f/e)$, is proportional to the mean value of
$-\ln(f/e)$. Clearly the entropy (\ref{CA6}), in the classical limit
reduces to the Boltzmann entropy. The constant $\gamma$, given by
(\ref{gamma}), represents the Lorentz factor related to the velocity
$v_*$ appearing in (\ref{DA7}), and in the classical limit,
approaches the unity. On the other hand the constant $\epsilon$
given by (\ref{EPS}), represent a relativistic generalization of the
Napier number $e$, which recovers in the classical limit.

The entropy (\ref{CA6}) can be written explicitly as follows
\begin{eqnarray}
S= \frac{1}{2\kappa} \int d^3p
\,\,\bigg(\frac{f^{1-\kappa}}{1-\kappa} -
\frac{f^{1+\kappa}}{1+\kappa}\bigg) \ , \label{C8}
\end{eqnarray}
while the related stationary distribution (\ref{B8}) assumes the
form
\begin{eqnarray}
f=\exp_{\kappa}\!\big(\!-\beta\,I+\beta\,\mu \,\big) \ , \ \ \ \ \ \
\ \ \ \label{C9}
\end{eqnarray}
and reduces to the Maxwell-Boltzmann distribution, in the classical
$\kappa \rightarrow 0$ limit.

The distribution (\ref{C9}), in the global rest frame where
$U_{\nu}=(c,0,0,0)$ and in absence of external forces i.e.
$A^{\nu}=0$, simplifies as
\begin{eqnarray}
f=\exp_{\kappa}\left(- \beta\,E+\beta\,\mu \right) \ , \ \ \ \ \ \ \
\ \ \label{C10}
\end{eqnarray}
$E$ being the relativistic kinetic energy. This distribution at low
energies ($E \rightarrow 0$) reduces to the classical
Maxwell-Boltzmann distribution i.e. $f\propto \exp (-
\beta \,E )$, while at relativistic energies ($E
\rightarrow +\infty$) presents power-law tails i.e.
\begin{eqnarray}
f\propto E^{-1/\kappa} \ , \ \ \ \ \ \ \ \ \ \label{C11}
\end{eqnarray}
in accordance with the experimental evidence. We recall that the
first experimental validation of the distribution (\ref{C10}),
concerns cosmic rays and has been considered in ref. \cite{PRE2002}.
Recently a computer validation of the same distribution, has been
considered in refs. \cite{Lapenta,Lapenta2009}, where the relaxation
in relativistic plasmas under wave-particle interaction, has been
simulated numerically.

Finally, after posing $\Lambda(f)=\ln_{\kappa}(f)$, in  Eq.
(\ref{B4}), the relativistic two-particle correlation function
assumes the form
\begin{equation}
f\otimes f_1 =\exp_{\kappa} \left( \,\ln_{\kappa}\! f  +
\ln_{\kappa}\!f_1 \, \right )  \ ,  \label{C11}
\end{equation}
which in the $\kappa \rightarrow 0$ classical limit, reduces to
$ff_1$, as dictated by the molecular chaos hypothesis. Consequently
the relationship (\ref{C11}) can be viewed as defining the
relativistic extension of the molecular chaos hypothesis.

\section{Conclusions}

Let us consider the probability distribution function
\begin{eqnarray}
f=\xi_1\exp_{\kappa}\!\big(\!-\xi_2\,\beta\,[E-\mu] \,\big) \ ,
\label{H1}
\end{eqnarray}
$\xi_1$ and $\xi_2$ being two arbitrary constants. In the expression
of the distribution function $E$ is the relativistic microscopic
energy, while $\beta$ and $\beta \mu$ are the Lagrange multipliers.
The Maximum Entropy Principle asserts that the distribution
(\ref{H1}) can be obtained,  by maximizing the entropy
\begin{eqnarray}
S=- \,\frac{\gamma}{\xi_2}\int d^3p \,\,\, f
\,\ln_{\kappa}(f/\,\epsilon \, \xi_1 ) \ , \label{H2}
\end{eqnarray}
with $\gamma=1/\sqrt{1-\kappa^2}$ and $\epsilon =
\exp_{\kappa}(\gamma)$, under the constraints imposing the
conservation of the norm of $f$ and of the mean value of the $E$. It
is remarkable the above distribution and entropy are linked through
the Maximum Entropy Principle, independently on the particular
values of the arbitrary constants $\xi_1$ and $\xi_2$.

In kinetic theory, customarily it is posed
\begin{eqnarray}
\xi_1=\xi_2=1 \ ,
\end{eqnarray}
in order to simplify the expression of the distribution function and
of the two particle correlation function, appearing in kinetic
equation. This choice, is made naturally, in the present paper,
being the starting point of our presentation the kinetic equation
(\ref{B1}).

On the other hand, in statistical mechanics, in order to simplify
the expression of the entropy, it is posed
\begin{eqnarray}
&& \xi_1=1/\epsilon \ ,\\
&& \xi_2=\gamma \ .
\end{eqnarray}
This latter choice has been made for instance in the ref. \cite{PRE2002,PRE2005}
where the starting point for the presentation of the theory was the
entropy functional.

\end{document}